\documentclass{mem}
\usepackage{amsmath}
\usepackage{natbib}
\usepackage{txfonts}
\usepackage{balance}
\usepackage{graphicx}
\usepackage{flushend}
\usepackage[breaklinks,pdftex]{hyperref}
\usepackage{amssymb}
\usepackage[T1]{fontenc}
\usepackage[utf8]{inputenc}
\usepackage{color} 
\idline{75}{149}

\newcommand{\g}{\ensuremath{\gamma}}
\newcommand{\E}[1]{\times 10^{#1}}

\begin{document}
\def\teff{$T\rm_{eff }$}
\def\kms{$\mathrm {km s}^{-1}$}

\title{
Multiwavelength correlation studies in the era of CTAO
}


\author{
Michael Zacharias\inst{1,2} 
\and Elina Lindfors\inst{3} 
\and Patrizia Romano\inst{4}
\and Daniela Dorner\inst{5} 
\and Stefano Vercellone\inst{4}
\and Matteo Cerruti\inst{6} 
\and Jonathan Biteau\inst{7,8} 
for the CTAO Consortium
          }

\institute{
Landessternwarte, Zentrum für Astronomie Heidelberg (ZAH), Universit\"{a}t Heidelberg, K\"{o}nigstuhl 12, D-69117 Heidelberg, Germany \email{m.zacharias@lsw.uni-heidelberg.de}
\and Centre for Space Research, North-West University, Potchefstroom 2520, South Africa \email{mzacharias.phys@gmail.com}
\\
The remaining affiliations can be found at the end of the paper.
}

\authorrunning{M.~Zacharias et al.}

\titlerunning{MWL correlations in the CTAO era}

\date{Received: Day Month Year; Accepted: Day Month Year}

\abstract{
Correlations between various multiwavelength (MWL) bands are an intermittent feature in blazar light curves; that is, they are observed in some instances but not in others. With the CTAO we will obtain detailed very-high-energy (VHE) gamma-ray light curves for many sources also during their low states, enabling detailed MWL correlation studies. For two blazars, the HBL Mrk\,421 and the FSRQ PKS\,1510-089, the long-term X-ray and optical light curves are used to induce variations in input parameters of the lepto-hadronic one-zone code OneHaLe. We show light curves in the CTA energy range for three different energy thresholds. The results are: 1) the presence of relativistic protons has a significant effect on the correlation of the light curves as the emerging pair cascade prolongs flaring states at the highest energies; and 2) comparison of the theoretical light curves with existing VHE gamma-ray data shows that both leptonic and hadronic models can only partially reproduce the data.
\keywords{Relativistic processes -- radiation mechanisms: non-thermal -- BL Lacertae objects: individual (Mrk\,421) -- Quasars: individual (PKS\,1510-089)}
}
\maketitle{}

\section{Introduction}

In blazars, the relativistic jet of the distant active galactic nucleus points towards the Earth \citep{urrypadovani95}. Blazars are known for their strong variability in all energy bands, as well as their double-humped spectral energy distribution (SED). The low-energy hump in the SED is due to synchrotron emission of relativistic electrons, while the nature of the high-energy hump is a matter of debate. In leptonic models, this is explained by inverse-Compton scattering of soft photon fields by the same electrons. In hadronic models, relativistic protons can produce \g\ rays directly through proton-synchrotron emission or indirectly through pion production and the corresponding cascade emission \citep{cerruti20}.

The multiwavelength (MWL) variability can range from months down to minutes time scales, and even ``quiescent'' states are marked by low-amplitude fluctuations. Simple models suggest that blazar MWL variability should be correlated between different energy bands. Therefore, a careful study of these correlation patterns can reveal the plausibility of these models.

Here, we invert this strategy: We employ the data of an observed synchrotron light curve, feed it into a radiation code and predict the \g-ray light curves that could become observable with the CTAO. Using various setups (leptonic, hadronic and lepto-hadronic), we highlight the observable differences between these setups.
As examples, we chose two well-known blazars: the high-frequency-peaked BL Lac object (HBL) Markarian~421, and the flat-spectrum radio quasar (FSRQ) PKS~1510-089. 

\section{Strategy}

In order to generate MWL light curves from the observed light curve in a specific band  (X-ray or optical), we need to create a template pattern, $P(t)$, that can be used as input to vary the particle density in a time-dependent radiation code. If we only consider the integrated flux changes from the observations, the pattern becomes
\begin{align}
	P(t) = \frac{F(t)-F_{\rm low}}{F_{\rm low}}
	\label{eq:pattern},
\end{align}
where $F_{\rm low}$ is the minimum flux in the light curve ensuring $P(t)\geq0$. 
The observations in the data sets are typically not equidistant in time. However, an equal time spacing is advantageous for the simulations. Thus, a linear interpolation is done between subsequent light curve points.

As stated, $P(t)$ is used to create a time-dependent variation in the particle density in the code. This takes the form
\begin{align}
	n(t) = n_0 [1+P(t)]
	\label{eq:densvar},
\end{align}
with the initial density $n_0$ recreating the minimum flux of the light curve.

In this work, we use the time-dependent, lepto-hadronic one-zone code \texttt{OneHaLe} \citep{zacharias21,zacharias+22}. The \g-ray flux is produced with three different realizations: (1) a leptonic setup with inverse-Compton scattering of synchrotron and/or external fields, (2) a hadronic setup with proton-synchrotron as the main \g-ray radiation process, and (3) a lepto-hadronic setup, where the leptonic case is augmented  by a population of relativistic protons causing a pair cascade that in turn emits synchrotron \g\ rays. The baseline or low-state parameters for each model are given in Tab.~\ref{tab:params}. We note that absorption of \g\ rays in the extragalactic background light is not accounted for.

\begin{table*}
\begin{footnotesize}
\caption{List of low-state parameters for Mrk\,421 and PKS\,1510-089.}
\label{tab:params}
\begin{center}
\vspace{-0.5cm}
\begin{tabular}{l|ccc} 
	\hline
	\textbf{Source} & \multicolumn{3}{c}{\textbf{Mrk\,421}}  \\
	Model	& leptonic & hadronic & lepto-hadronic  \\
	\hline
	Mag. field & $0.17\,$G & $10\,$G & $0.17\,$G  \\
	Radius & $1.5\E{16}\,$cm & $1.5\E{16}\,$cm & $1.5\E{16}\,$cm  \\
	Doppler factor & $25$ & $25$ & $25$  \\
	e inj. luminosity & $1.5\E{40}\,$erg/s & $1\E{40}\,$erg/s & $1.5\E{40}\,$erg/s  \\
	e $\gamma_{\rm min}$ / $\gamma_{\rm max}$ & $2.2\E{4}$ / $4\E{5}$ & $3\E{3}$ / $4\E{4}$ & $2.2\E{4}$ / $4\E{5}$  \\
	p inj. luminosity & --- & $7\E{40}\,$erg/s & $3\E{42}\,$erg/s   \\
	p $\gamma_{\rm min}$ / $\gamma_{\rm max}$ & --- / --- & $2$ / $1\times 10^{10}$ & $2$ / $1\times 10^{8}$  \\
	e / p spectral index & $3.5$ / --- & $3.5$ / $2.0$ & $3.5$ / $1.5$  \\
	Escape time & $35\,R/c$ & $35\,R/c$ & $35\,R/c$  \\
	\hline
	\hline
	\textbf{Source} & \multicolumn{3}{c}{\textbf{PKS\,1510-089}} \\
	Model	& leptonic & hadronic & lepto-hadronic  \\
	\hline
	Mag. field & $0.12\,$G & $30\,$G & $0.12\,$G \\
	Radius & $1.0\E{16}\,$cm & $6.0\E{16}\,$cm & $1.0\E{16}\,$cm \\
	Doppler factor & $20$ & $20$ & $20$ \\
	e inj. luminosity & $1.3\E{42}\,$erg/s & $3.5\E{40}\,$erg/s & $1.3\E{42}\,$erg/s \\
	e $\gamma_{\rm min}$ / $\gamma_{\rm max}$ & $1.5\E{3}$ / $3\E{5}$ & $1.1\E{2}$ / $4\E{4}$ & $1.5\E{3}$ / $5\E{4}$ \\
	p inj. luminosity & --- & $1\E{43}\,$erg/s & $3\E{43}\,$erg/s  \\
	p $\gamma_{\rm min}$ / $\gamma_{\rm max}$ & --- / --- & $2$ / $5\times 10^{8}$ & $2$ / $1\times 10^{8}$  \\
	e / p spectral index & $3.0$ / --- & $3.0$ / $1.9$ & $3.0$ / $1.5$ \\
	Escape time & $10\,R/c$ & $10\,R/c$ & $10\,R/c$ 
\end{tabular}
\vspace{-0.5cm}
\end{center}
\end{footnotesize}
\end{table*}

\section{Markarian 421}

\begin{figure*}[htb]
\resizebox{\hsize}{!}{
\includegraphics[width=0.49\textwidth]{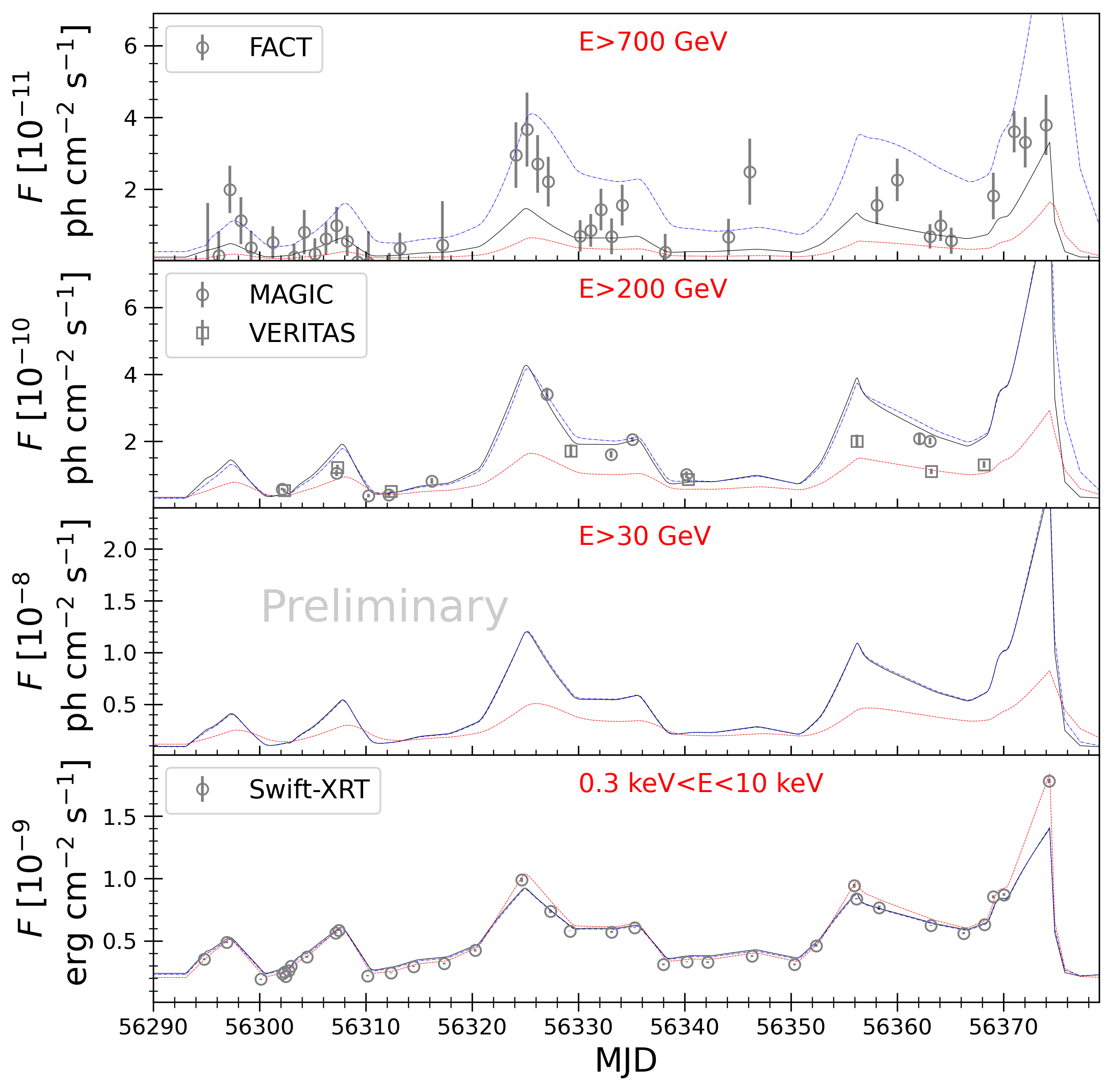}
~
\includegraphics[width=0.49\textwidth]{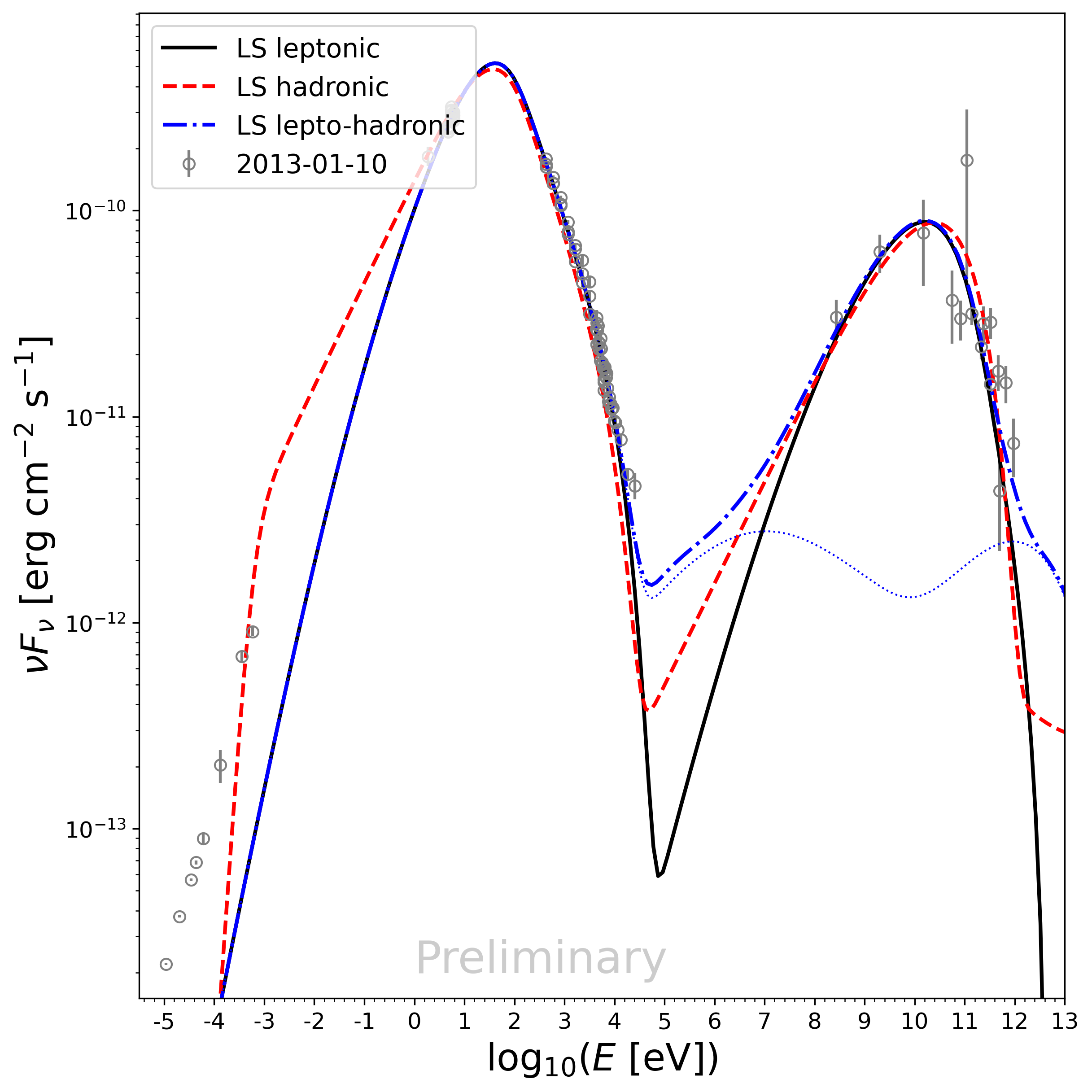}
}
\caption{\footnotesize
\textbf{Left:} Simulated light curves for Mrk\,421 for the models and energy ranges as labeled. Data from \cite{MB+16} and \cite{AA+21}. \textbf{Right:} Low-state spectra for the three models. The blue dotted line shows the electron-synchrotron flux including pairs. Data from \cite{MB+16}.
}
\label{fig:mrk421}
\end{figure*}

Mrk\,421 is an HBL at redshift $z=0.0308$. Its proximity makes it one of the most observed blazars from the ground and from space across all energies. In most circumstances, its spectrum is described well with a steady-state, one-zone SSC model \citep{MB+16,AA+21}. 
The low-state spectra are shown in Fig.~\ref{fig:mrk421} (right), and all three realizations are capable of reproducing the data.

We use the X-ray light curve of a ``low state'' in early 2013 \cite{MB+16} to derive the variability pattern, $P(t)$. This is used to vary the electron and proton densities (the latter only if applicable) resulting in the light curves shown in Fig.~\ref{fig:mrk421} (left). The X-ray light curve -- which, again, served as the variability input -- is reproduced well except for the two brightest flux states in the leptonic and lepto-hadronic model. This may have to do with the choice to only vary the electron normalization but not the spectral index. The latter can have significant influence on the integrated flux if the spectrum were to become much harder. 

Figure~\ref{fig:mrk421} (left) also shows three \g-ray light curves relevant for the CTAO energy range. 
It highlights the importance of the chosen energy threshold to distinguish between the different scenarios.
Most notably, the hadronic setup produces the least variability, as the proton-synchrotron flux only depends linearly on the proton density. In the SSC-dominated cases, the \g-ray flux depends on the square of the electron density, inducing stronger variability. Interestingly, the comparison to the \g-ray data collected with MAGIC, VERITAS and FACT provides strong evidence that the simple setup chosen here is not sufficient to reproduce the data. While the light curve for the threshold of $200\,$GeV is mostly reproduced well, there are clear differences in the FACT domain at $E>700\,$GeV. Here, the contribution from the pair cascade becomes evident and seems to be capable of improving the fit to the data in several cases compared to the simple SSC scenario. On the other hand, the hadronic setup is not capable of reproducing any of the \g-ray data.


\section{PKS\,1510-089}

\begin{figure*}[htb]
\resizebox{\hsize}{!}{
\includegraphics[width=0.49\textwidth]{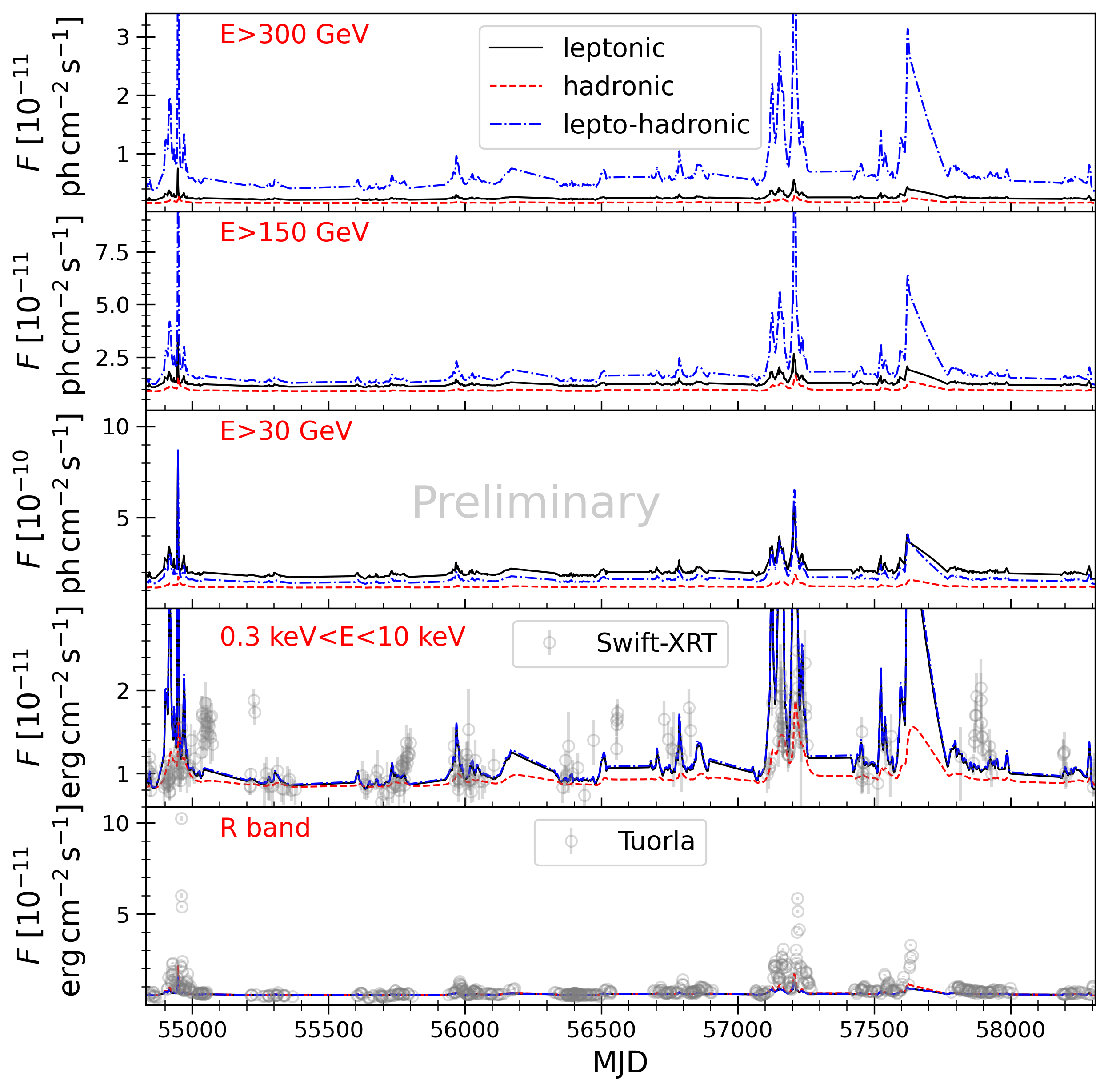}
~
\includegraphics[width=0.49\textwidth]{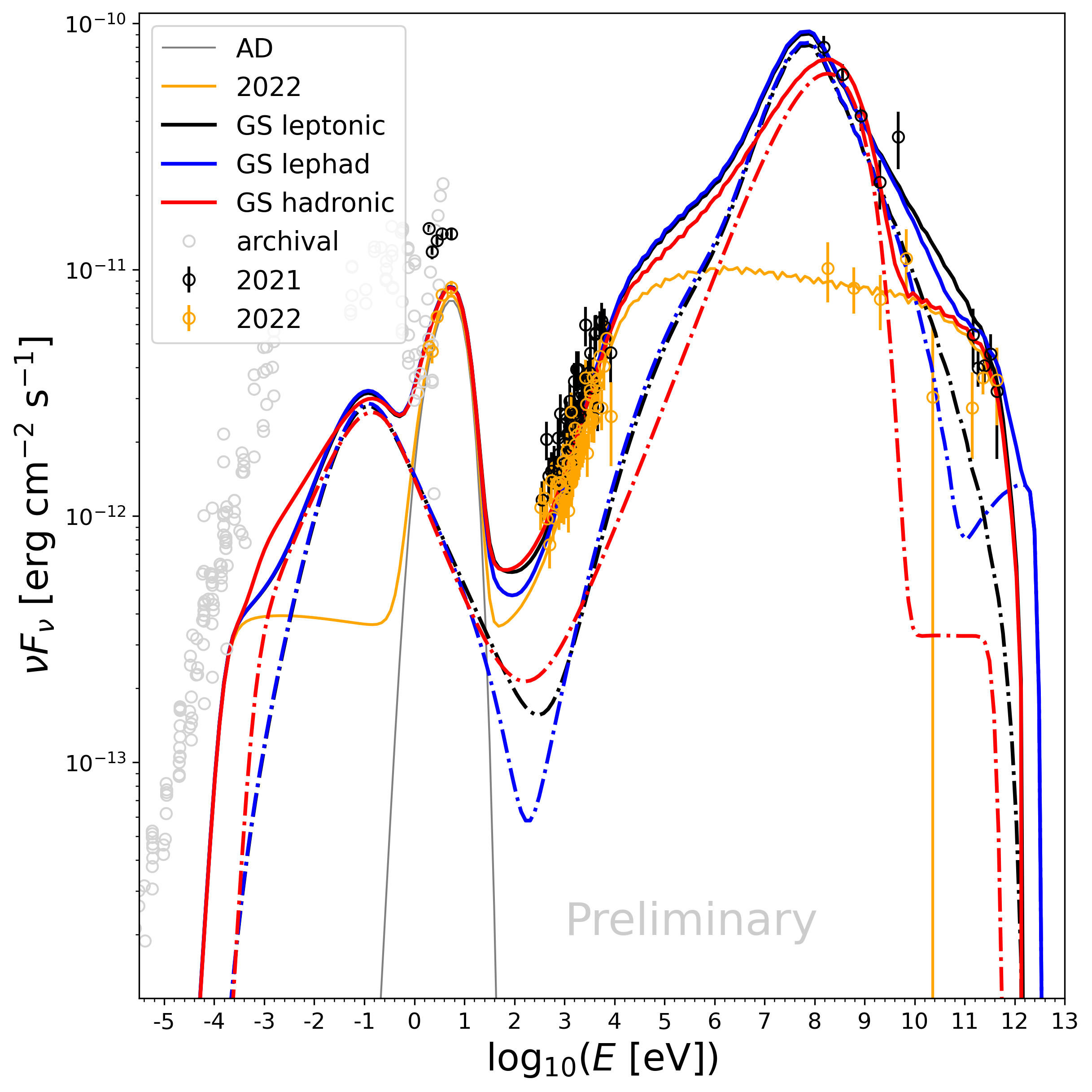}
}
\caption{\footnotesize
\textbf{Left:} Simulated light curves for PKS~1510-089 for the models and energy ranges as labeled. Data from \cite{KN+18} and \textit{Swift} (this work). \textbf{Right:} Low-state two-zone models as labeled (total model: solid; variable zone: dot-dashed). Data from \cite{hess23}.
}
\label{fig:pks1510}
\end{figure*}

PKS\,1510-089 is an FSRQ at redshift $z=0.361$. It is known for its bright external photon fields, with a prominent accretion disk (big blue bump) and equally prominent broad emission lines. For a long time, it was thus speculated that the \g\ rays should be produced by inverse-Compton scattering of the broad-line region (BLR). However, the continuous detection of PKS~1510-089 at VHE \g\ rays \cite{hess13,magic18,zacharias+19,hess+21} challenges this interpretation, as normally the VHE photons should be absorbed in the bright BLR \cite{meyerscarglebladford19}.

Even more astonishing, PKS\,1510-089 underwent a massive change of character in July 2021 \citep{hess23}, when the HE \g-ray flux dropped significantly, and the optical polarization vanished. Surprisingly, the VHE \g-ray flux did not change. This lead \cite{hess23} to conclude that PKS~1510-089 used to contain at least two active emission regions, from which an inner (and maybe normally more active) region vanished in 2021 leaving behind only the outer, VHE \g-ray producing zone. We follow this two-zone approach here. That is, we use a steady outer zone to produce a baseline flux in the X-ray and VHE \g-ray domain and add a variable inner zone that produces the optical and HE \g-ray flux observed pre-2021.

The external photon fields (accretion disk and dusty torus) are described as in \cite{hess23}.
In order to produce the synchrotron-based variability pattern, $P(t)$, the optical R-band light curve is used \citep{KN+18}. The light curves and low state spectra are shown in Fig.~\ref{fig:pks1510} (left and right), respectively.

It is obvious from Fig.~\ref{fig:pks1510} (left) that the variability in the R band cannot be reproduced. We suspect that the bright accretion disk, which requires a rather low synchrotron flux during low states, is too dominant. That is, the synchrotron flux would have to rise significantly to contribute in the optical domain. 

The X-ray band data are also not reproduced well. In many cases, the model X-ray variability is much larger than the observed one. Additionally, several X-ray high states are not covered by the model -- in many cases likely because of non-simultaneity of the data.

The three exemplary \g-ray bands typically show similar variability patterns, as the inverse-Compton flux using external photons depends linearly on the electron density, as does the proton-synchrotron flux. However, the flux ratios between the different bands differ depending on the setup (most notably in the lepto-hadronic component due to the pair-cascade contribution), which may be an additional tool to distinguish between different models with future data.

In any case, it is clear that these simple setups are incapable of reproducing the long-term trend of PKS~1510-089. A more complicated approach is, however, beyond the scope of this contribution.

\section{Summary}

Both sources suggest that a simple scaling by particle density does not reproduce well the high-energy component.
Mrk\,421 is known to adhere to simple SSC models during flares, but our model suggests that this is not the case for longer periods of time (however, spectral changes in the X-ray domain have not been accounted for).
PKS\,1510-089 is a much more complicated source, which was recently found to probably have at least two emission zones.
It was not possible to properly reproduce either the optical light curve or the complicated optical spectrum with our simple scaling model.
Nevertheless, flux ratios between various \g-ray light curves may help distinguishing between scenarios.

CTAO's superior capabilities to current-generation instruments especially at energies far above the peak energy will much more strongly constrain the models, allowing for more complicated setups (beyond mere density variations) to be tested. 

\vspace{0.2cm}


\noindent {\bf{Affiliations}}\par
{\footnotesize
$^{3}$ Department of Physics and Astronomy, University of Turku, Finland\par
$^{4}$ INAF - Osservatorio Astronomico di Brera, Via E. Bianchi 46, 23807 Merate (LC), Italy\par
$^{5}$ Julius-Maximilians-Universität Würzburg, Fakultät für Physik und Astronomie, Institut für Theoretische Physik und Astrophysik, Lehrstuhl für Astronomie, Emil-Fischer-Str. 31, D-97074 Würzburg, Germany\par
$^{6}$ Universit\'{e} Paris Cit\'{e}, CNRS, Astroparticule et Cosmologie, F-75013 Paris, France\par
$^{7}$ Universit\'e Paris-Saclay, CNRS/IN2P3, IJCLab, 91405 Orsay, France\par
$^{8}$ Institut universitaire de France (IUF), France
}

\begin{acknowledgements}
This work was conducted in the context of the CTAO Consortium. We gratefully acknowledge financial support from the agencies and organizations listed here:\\ \url{https://www.ctao.org/for-scientists/library/acknowledgments/}.\\
MZ acknowledges funding by the Deutsche Forschungsgemeinschaft (DFG, German Research Foundation) project number 460248186 (PUNCH4NFDI). The \texttt{OneHaLe} code is available upon reasonable request to M.~Zacharias.
\end{acknowledgements}
\vspace{-0.5cm}

\bibliographystyle{aa}
\bibliography{bibliography}

\end{document}